\documentclass[conference]{IEEEtran}
\IEEEoverridecommandlockouts
\usepackage{amsmath,graphicx}
\usepackage{hyperref}
\usepackage{amssymb}
\usepackage{CJKutf8}
\usepackage{algorithm}
\usepackage{algorithmic}
\usepackage{multirow}
\usepackage{siunitx}
\usepackage{booktabs}
\usepackage{amssymb}

\usepackage{cite}
\usepackage{amsmath,amssymb,amsfonts}
\usepackage{algorithmic}
\usepackage{graphicx}
\usepackage{textcomp}
\usepackage{xcolor}
\def\BibTeX{{\rm B\kern-.05em{\sc i\kern-.025em b}\kern-.08em
    T\kern-.1667em\lower.7ex\hbox{E}\kern-.125emX}}
\begin{document}

\title{Developing a Multilingual Dataset and Evaluation Metrics for Code-Switching: A Focus on Hong Kong's Polylingual Dynamics\\

\thanks{Funding agency: HongKongRGC 16308221, 16310222, HongKong ITF PRP/077/20FX, PRP/055/23FX}
}

\author{\IEEEauthorblockN{Peng Xie}
\IEEEauthorblockA{\textit{
Department of Computer Science and Engineering} \\
\textit{The Hong Kong University of Science and Technology}\\
Hong Kong, China \\
pxieaf@connect.ust.hk}
\and
\IEEEauthorblockN{Kani Chen}
\IEEEauthorblockA{\textit{
Department of Mathematics} \\
\textit{The Hong Kong University of Science and Technology}\\
Hong Kong, China \\
makchen@ust.hk}
}

\maketitle

\begin{abstract}
The existing audio datasets are predominantly tailored towards single languages, overlooking the complex linguistic behaviors of multilingual communities that engage in code-switching. This practice, where individuals frequently mix two or more languages in their daily interactions, is particularly prevalent in multilingual regions such as Hong Kong, China. To bridge this gap, we have developed a 34.8-hour dataset of Mixed Cantonese and English (MCE\footnote{MCE Dataset and finetuned model weight are available at \url{https://github.com/Shelton1013/Whisper\_MCE}}) audio using our Multi-Agent Data Generation Framework (MADGF). We fine-tuned the open-source multilingual Automatic Speech Recognition (ASR) model, Whisper, with the MCE dataset, leading to impressive zero-shot performance. The traditional metrics overlook important factors such as latency in real-world applications and code-switching scenarios. We have introduced a novel evaluation metric called Fidelity to the Original Audio, Accuracy, and Latency (FAL). This metric aims to overcome the limitations of traditional metrics used to assess ASR systems.
\end{abstract}


\section{Introduction}
In the domain of natural language processing and machine learning, the advancement and refinement of audio recognition technologies heavily depend on the accessibility and quality of audio datasets. Historically, these datasets have primarily comprised monolingual audio recordings, employed for training models to execute tasks such as speech recognition. The Hong Kong Cantonese Adult Language Corpus (HKCAC) \cite{leung2001hkcac} has 8.1 hours of transcripts and approximately 170,000 characters. The Hong Kong Cantonese Corpus (HK-CanCor) consists of 30.0 hours of recordings, with each sample 10 minutes long. The Corpus of Mid-20th Century Hong Kong Cantonese (HKCC) has in total about 200,000 character tokens. The Hong Kong Cantonese MapTask Corpus (CantoMap) \cite{winterstein2020cantomap} contains a total of 768 minutes of recordings and transcripts of forty speakers. The Common Voice zh-HK \cite{ardila2019common} is a massive multilingual collection of transcribed speech collected and validated via Mozilla’s Common Voice initiative. The Multi-Domain Cantonese Corpus (MDCC) \cite{yu2022automatic} consists of 73.6 hours of clean read speech paired with transcripts, collected from Cantonese audiobooks from Hong Kong. The details of the dataset are enumerated in Table \ref{Cantonese ASR Corpora}. 

Code-switching \cite{gardner2009code} is the practice of alternating between two or more languages within a single conversation. This linguistic behavior is widespread in multilingual societies, particularly in settings where different languages frequently come into contact. People may switch languages for a variety of social, cultural, psychological, or linguistic purposes. For example, in Hong Kong\cite{li2000cantonese}, individuals often use both English and Cantonese in the same sentence, particularly in business and educational contexts. Similarly, within the Hispanic community in the United States, it is common for people to transition between English and Spanish\cite{li2008understanding} in both domestic and communal environments.

\begin{figure*}
  \centering
  \includegraphics[width=0.7\textwidth]{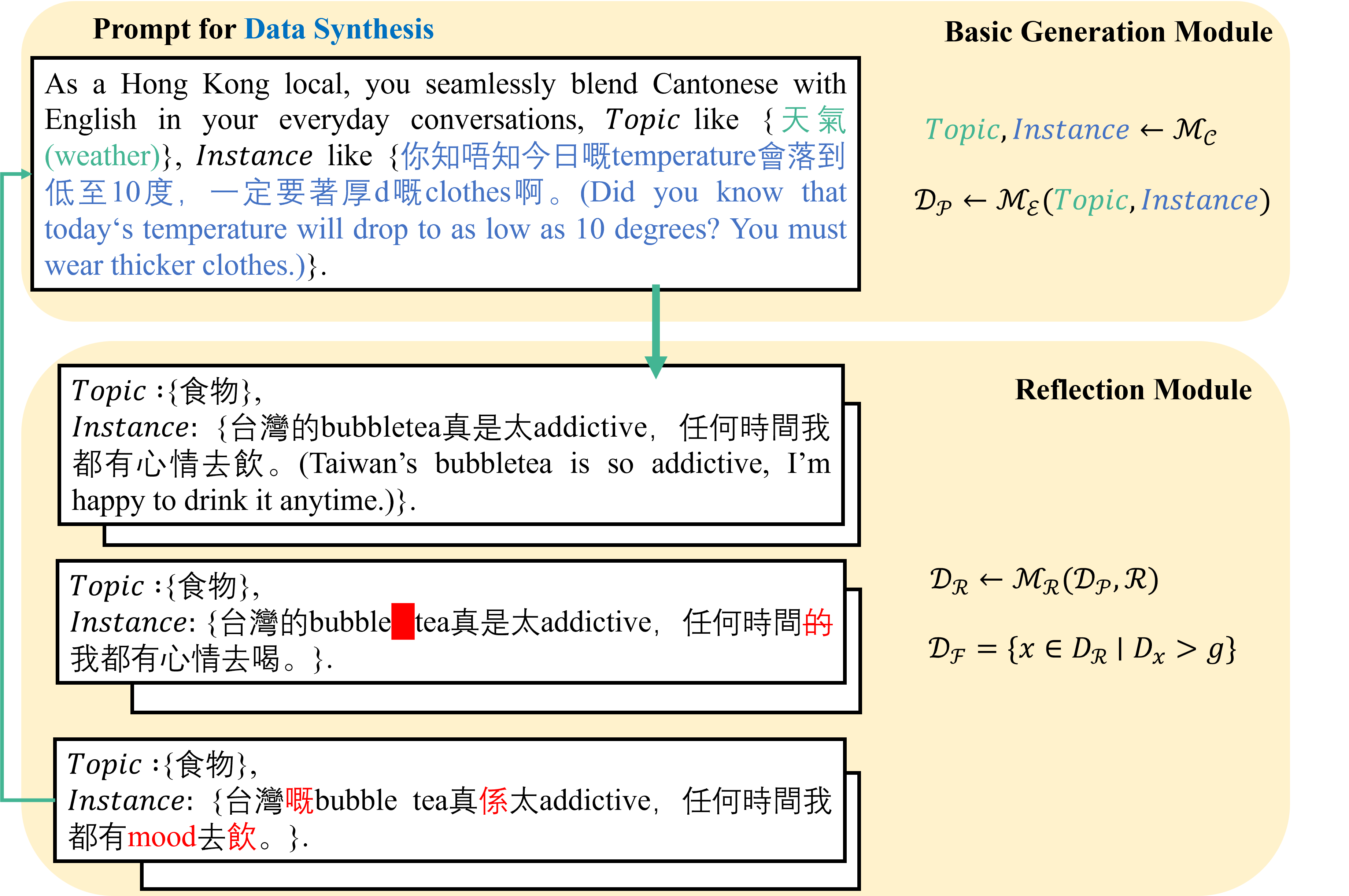}
  \caption{\textbf{Overview of MADGF.} MADGF consists of two modules that integrate multiple specialized agents to cooperate in generating text data. The Basic Generation Module utilizes a Creator Agent and an Engineer Agent to generate fundamental data. The Reflection module employs a Reflector Agent to check the grammar and enhance the data. \\}
  \label{Pipeline}
\end{figure*}

However, the available datasets for Cantonese are not necessarily representative of how people typically speak in Hong Kong. In daily life, local people always speak Cantonese mixed with English, but there is a lack of datasets that reflect this reality. Moreover, the existing datasets are of low quality and may contain errors, such as mislabeled recordings. For instance, the MDCC\cite{yu2022automatic} dataset uses \{\} symbols to indicate uncertainty about certain words. Therefore, we propose MADGF to generate a high-quality, diverse dataset, as shown in Figure \ref{Pipeline}.

End-to-end (E2E) models\cite{li2004analysis} have shown impressive performance in ASR accuracy benchmarks. However, they have practical limitations\cite{wang2003word} when it comes to streaming, latency, and code-switching scenarios, which hinder their widespread commercial adoption. Speech recognition and machine translation systems are typically evaluated using metrics like Word Error Rate (WER), Character Error Rate (CER), and Mixed Error Rate (MER)\cite{kang2011mandarin}. However, these metrics may be too strict for certain applications, especially in code-switching scenarios, as they do not account for contextual or stylistic nuances such as font variations or script differences. Additionally, these metrics only focus on accuracy and do not consider the important aspect of latency in real-world applications.

Our major contributions can be summarized as follows:
\begin{itemize}
\item We introduce a Multi-Agent Data Generation Framework, MADGF, which utilizes multiple agents to produce high-quality datasets.
\item We publish a unique 34.8-hour high-quality mixed language audio dataset named MCE, which covers a broad spectrum of topics and provides rich code-switching scenarios.
\item We propose a novel ASR evaluation metric FAL, which takes into consideration three pivotal metrics: Fidelity to the Original Audio, Accuracy, and Latency. The inclusion of these criteria makes FAL an essential tool for developers and researchers aiming to refine ASR technologies to better meet real-world demands and enhance user experience.
\end{itemize}

\section{CORPUS DESIGN}
\subsection{MADGF}
When using a large language model to generate data, it heavily relies on human-written instruction data. The use of a simple and fixed prompt may lead to limitations in terms of quantity, diversity, and creativity \cite{wang2022self}. Due to the limited availability of high-quality open-source mixed Cantonese and English audio datasets, we utilized MADGF to create such a dataset.

\begin{table}[htbp]
  \caption{Cantonese ASR Corpora}
  \centering
  \begin{tabular}{cccc}
    \hline
    Name & Data source & Code-switching \\
    \hline
    HKCAC & Phone-in programs & $\times$ \\
    HKCanCor& Chat  & $\times$\\
    HKCC& Movie  & $\times$ \\
    Common Voice zh-HK & MapTask  & $\times$\\
    CantoMap& Wikipedia  & $\times$\\
    MDCC & Audiobook & $\times$\\
    \hline
    \textbf{MCE(ours)} & \textbf{LLM} & \textbf{\checkmark}\\
    \hline
  \end{tabular}
  \label{Cantonese ASR Corpora}
\end{table}

MADGF is an innovative multi-agent system designed to produce high-quality, mixed-language datasets. This framework breaks complex tasks into subtasks \cite{shen2024hugginggpt} and integrates multiple specialized agents \cite{wu2023autogen}, each contributing uniquely towards the creation and refinement of the dataset. Our architecture allows for the efficient handling of diverse data types and enhances the authenticity and applicability of the generated datasets for real-world applications.

The Creator Agent $\mathcal{M}_{\mathcal{C}}$\cite{zhang2022bigssl} is responsible for defining the scope and specifics of the data to be generated. This role is crucial as it sets the Topic and Instance, ensuring that the generated data are relevant and targeted. 

\begin{equation}
    \textbf{Topic}, \textbf{Instance} \leftarrow \mathcal{M}_{\mathcal{C}}
\end{equation}

The Engineer Agent $\mathcal{M}_{\mathcal{E}}$ crafts detailed prompts that guide the data generation process. 

\begin{equation}
    \mathcal{D}_{\mathcal{P}} \leftarrow \mathcal{M}_{\mathcal{E}}(\textbf{Topic},\textbf{Instance})
\end{equation}

The Reflector Agent $\mathcal{M}_{\mathcal{R}}$ enhances and evaluates the output. This role involves critical reflection $\mathcal{R}$\cite{park2023generative} on the generated data and $g$ represents the grade threshold used for filtering the data. This feedback loop is vital as it informs the Prompt Engineer on how to adjust the prompts to optimize future data generation.

\begin{align}
    \mathcal{D}_{\mathcal{R}} &\leftarrow \mathcal{M}_{\mathcal{R}}(\mathcal{D}_{\mathcal{P}},\mathcal{R})\\
    \mathcal{D}_{\mathcal{F}} &= \{x\in\mathcal{D}_{\mathcal{R}} | \mathcal{D}_{x}>g\}
\end{align}

\begin{figure}
  \centering
  \includegraphics[width=0.5\textwidth]{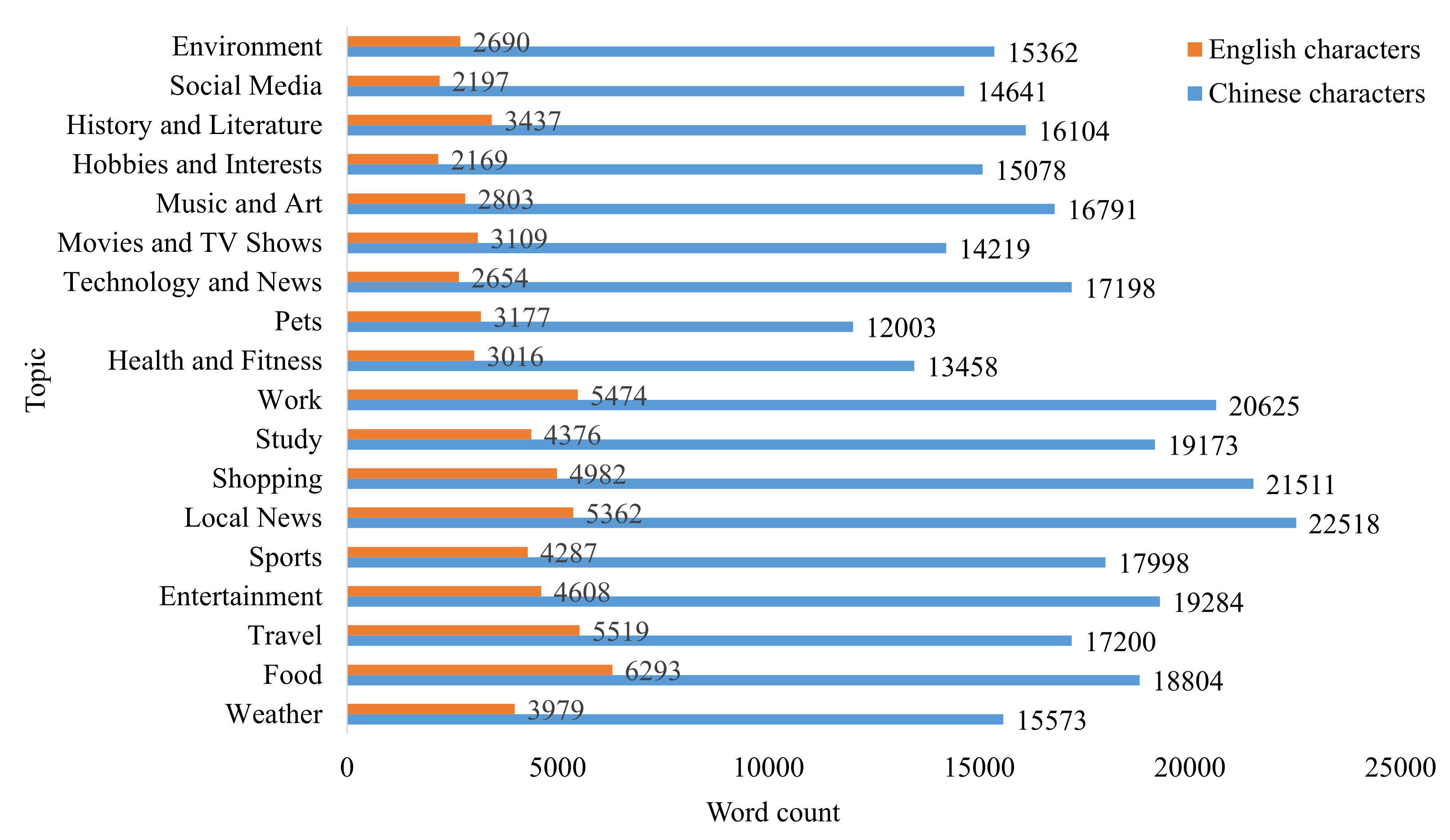}
  \caption{ MCE text dataset composition. The MCE dataset covers 18 daily topics and comprises over 16,000 sentences, with each sentence containing a mixture of Traditional Chinese and English characters. \\}
  \label{fig:MCE_text}
\end{figure}

\subsection{FAL}
The WER, CER, and MER metrics may be too stringent for certain applications, especially in code-switching scenarios, as they do not consider contextual or stylistic nuances such as font variations or script differences. Instead, we rely on Fidelity to the Original Audio $\mathcal{F}$ to evaluate code-switching or script differences, which are complex and challenging to assess using a specific formula. In these cases, subjective judgment plays a significant role. Many researchers use GPT4\cite{zheng2023judging} to evaluate the performance of other large language models. Therefore, we use GPT4 to calculate the Fidelity to the Original Audio.

Additionally, these metrics solely focus on accuracy\cite{kuchaiev2019nemo} and do not take into account the crucial aspect of Latency $\mathcal{L}$\cite{yu2020dual} in real-world applications. As a result, we have developed a novel evaluation metric called FAL to assess the performance of a speech-to-text system. FAL can be computed as:

\begin{scriptsize}
\begin{equation}
    \text{FAL} = \alpha \mathcal{F} + \beta(1 - \frac{S_{m} + I_{m} + D_{m}}{N_{m}})\cdot 100 + \gamma \left(1 + \frac{\mathcal{L}-1}{M-1}\cdot(100-1)\right)
\end{equation}
\end{scriptsize}

where $I_{m}$ is the number of English word and Chinese character insertions, $D_{m}$ is the number of English word and Chinese character deletions, $S_{m}$ is the number of English word and Chinese character substitutions, $N_{m}$ is the number of English words and Chinese characters in the reference, and $M$ is the max latency. 

Considering the various application scenarios, we have assigned different weights, denoted as $\alpha$, $\beta$, and $\gamma$, to each part. For instance, in real-time translation scenarios\cite{chen2021developing}, where there is a high demand for quick responses, $\gamma$ should be assigned a larger value. In certain digital human application scenarios, where the overall meaning of the conversation is more important than word-for-word consistency, a smaller value can be assigned to $\beta$.

Higher FAL indicates better performance, a proficient ASR evaluation metric is expected to accurately capture the substantive content and semantic essence of the original audio, produce precise transcriptions, and exhibit prompt response times.

\section{STATISTICS OF MCE DATASET}

As shown in Figure \ref{fig:MCE_text}, our MCE dataset covers 18 topics related to daily life, comprising a total of 34.8 hours of audio files. The corresponding annotated text consists of 307,540 Chinese characters and 70,132 English words. Among the topics, the "Food" category has the highest frequency of English words, with a Chinese character to English word ratio of approximately 3:1. On the other hand, the "Tech News" topic has the lowest frequency of English words, approximately 8:1. We randomly sampled all audio files and divided them into training and testing sets in a 9:1 ratio. The resulting training set contains 31.3 hours of speech files, and the distribution of topics in the training and testing sets is relatively consistent.

\begin{figure}
  \centering
  \includegraphics[width=0.5\textwidth]{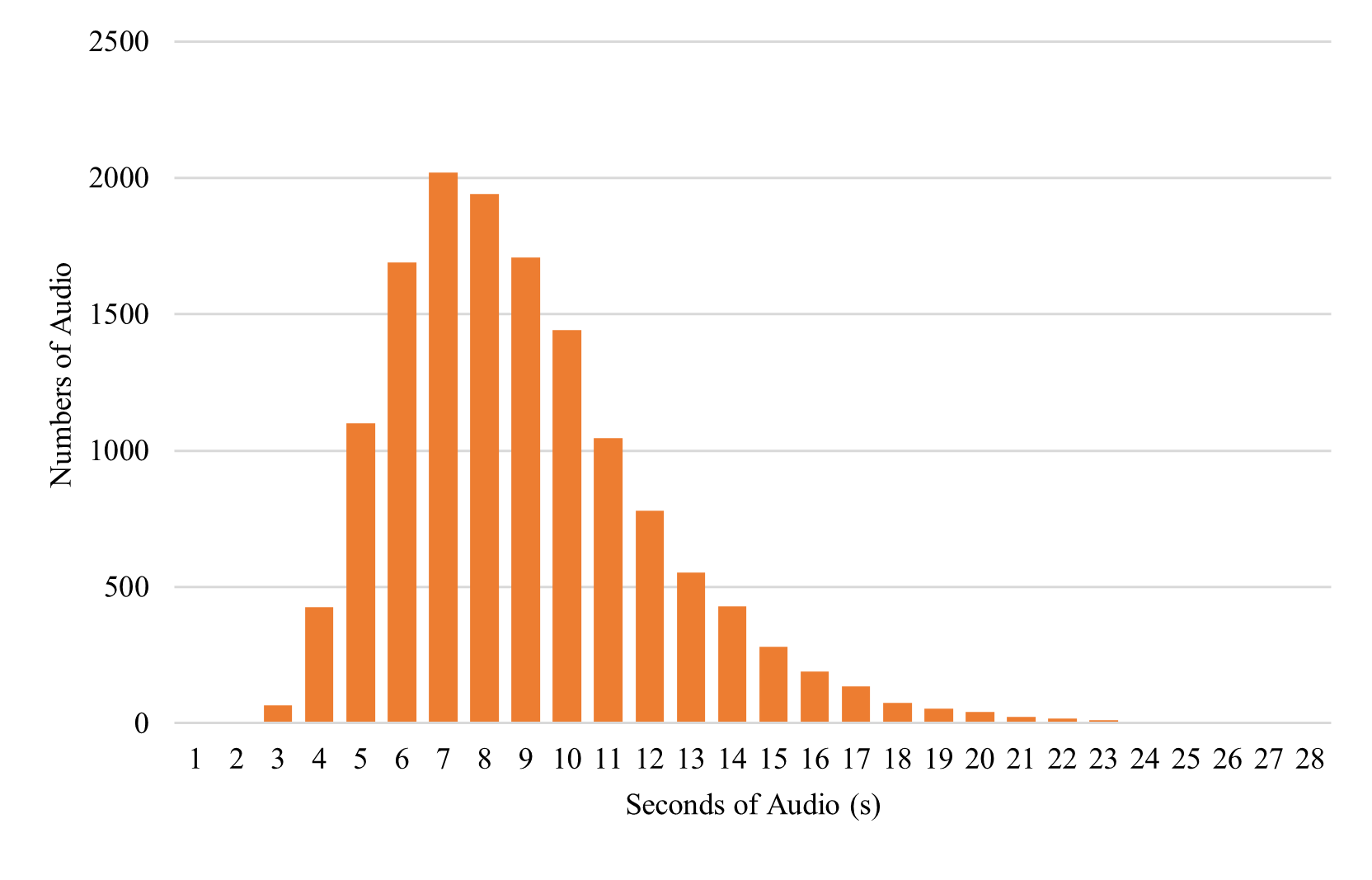}
  \caption{MCE audio dataset overivew.}
  \label{fig:MCE_audio}
\end{figure}

As shown in Figure \ref{fig:MCE_audio} The duration of audio files is concentrated in the 5-12 seconds range, with the longest audio file being 28 seconds. In most large-scale speech recognition models, there is no need for additional audio segmentation processing. During the audio recording, all volunteers replicated their habitual speaking speed, intonation, and other speaking habits from daily life. Volunteers with both fast and slow speech rates were selected, with faster speech rates potentially presenting more challenges for accurate recognition due to increased assimilation or pronunciation inaccuracies.

\begin{table*}
\caption{The fine-tuned and original models' ability to handle code-switching problems and single-language speech recognition.}
\resizebox{\textwidth}{!}{%
\sisetup{
table-number-alignment = center,
table-figures-integer = 2,
table-figures-decimal = 2
}
\centering
\renewcommand\footnoterule{\kern -1ex}
\renewcommand{\arraystretch}{1.3}
    \begin{tabular}{l*{9}{S}}
    \hline
    \multirow{3}{*}{Model Parameter} & \multicolumn{4}{c}{MCE} & \multicolumn{4}{c}{Common Voice zh-HK} \\
    \cmidrule(r){2-5}\cmidrule(lr){6-9}   
    & \multicolumn{2}{c}{MER(\%)} & \multicolumn{2}{c}{FAL} & \multicolumn{2}{c}{CER(\%)} & \multicolumn{2}{c}{FAL} \\
    \cmidrule(r){2-3} \cmidrule(r){4-5} \cmidrule(r){6-7} \cmidrule(lr){8-9}
    & \multicolumn{1}{c}{Whisper-MCE} & \multicolumn{1}{c}{Whisper} & \multicolumn{1}{c}{Whisper-MCE} & \multicolumn{1}{c}{Whisper} & \multicolumn{1}{c}{Whisper-MCE} & \multicolumn{1}{c}{Whisper} & \multicolumn{1}{c}{Whisper-MCE} & \multicolumn{1}{c}{Whisper} \\
    \hline
    Tiny(39M) & 25.05 & 76.03 & 80.55 & 50.39 & 27.06 & 52.41 & 80.08 & 63.48\\
    Base(79M) & 20.85 & 61.93 & 84.54 & 58.72 & 19.08 & 44.95 & 81.28 & 68.33\\
    Small(244M) & \textbf{14.28} & 49.41 & \textbf{90.91} & 67.67 & 17.64 & 29.93 & \textbf{87.71} & 78.18\\
    Medium(769M) & 19.51 & 54.81 & 79.14 & 55.06 & 13.12 & 23.22 & 86.66 & 74.44\\
    Large(1550M) & 15.96 & 52.59 & 76.81 & 53.22 & \textbf{12.61} & 26.83 & 78.72 & 67.56\\
    \hline
    \end{tabular}
    }
    \label{mix and single}
\end{table*}

\begin{CJK*}{UTF8}{bsmi}
\begin{table*}[]
\caption{Performance of the fine-tuned and original models on code-switching.}
\resizebox{\textwidth}{!}{%
\renewcommand{\arraystretch}{1.5}
\begin{tabular}{ccc}
\hline
Model &
  Audio Information &
  Speech Recognition Result \\ \hline
Whisper-Large-v2 &
  \multirow{3}{*}{\begin{tabular}[c]{@{}c@{}}士多啤梨草莓来源于strawberry, 士多商店店舖来源于store, \\ 波恤球衣来源于ball shirt, 貼士提示来源于tips, \\ 梳化沙發来源于sofa.\\  Translation: 士多啤梨Strawberry derives from strawberry, 士多store derives from stores,\\ 波恤ball shirt derives from ball shirt, 貼士tip derives from tips,\\ 梳化sofa derives from sofa.\\ \end{tabular}} &
  \begin{tabular}[c]{@{}c@{}}草莓來源於strawberry,雙點店來源於store,\\ 球衣來源於ball shirt,提示來源於tips,\\ 沙發來源於sofa.\end{tabular} \\ \cline{3-3} 
Whisper-Small &
   &
  \begin{tabular}[c]{@{}c@{}}Sito Berry草莓来源于Sroberry, Sito Shop店店店来源于Store, \\ Ball shirt 球衣 来源于Ball shirt, Tipsy提示来源于Tips, \\Sauva沙发来源于Sauva.\end{tabular} \\ \cline{3-3} 
Whisper-MCE-Small &
   &
  \begin{tabular}[c]{@{}c@{}}士多啤梨草莓源於Strawberry, 士多商店店舖源於Store, \\ 波恤球衣源於Ball shirt, 貼士提示源於Tips,\\ 梳化沙發源於Sofa.\end{tabular} \\ \hline
\end{tabular}%
}
       \label{test result}
\end{table*}
\end{CJK*}

\begin{table}
\caption{Comparative Performance of ASR Models on Code-Switching and Single-Language Speech Recognition.\\}
\centering
\arrayrulewidth=0.005mm
\renewcommand\footnoterule{\kern -1ex}
\renewcommand{\arraystretch}{1.3}
    \begin{tabular}{l*{11}{S}}
    \hline
    \multirow{2}{*}{Model} & \multicolumn{2}{c}{MCE} & \multicolumn{2}{c}{Common Voice zh-HK} \\
    \cmidrule(r){2-3}\cmidrule(lr){4-5}
    & \multicolumn{1}{c}{MER(\%)} & \multicolumn{1}{c}{FAL} & \multicolumn{1}{c}{CER(\%)} & \multicolumn{1}{c}{FAL} \\
    \hline
    XLSR-53 &\text{69.8}&\text{51.3}&\text{64.7}&\text{54.2}\\    Whisper&\text{49.4}&\text{67.7}&\text{29.9}&\text{78.2}\\
    Whisper-MCE&\textbf{14.3}&\textbf{90.9}&\textbf{12.6}&\textbf{87.7}\\
    \hline
    \end{tabular}
    \label{diff asr model}
\end{table}

\section{EXPERIMENTS AND RESULTS}
To highlight the significance of our MCE dataset and FAL metric, we fine-tuned the whisper\cite{radford2023robust} model on five different versions and conducted a comparative analysis with the original and other models. During the fine-tuning process, we do not use any data augmentation or text normalization tricks. Additionally, we evaluated the performance of our fine-tuned model on the common voice zh-HK to test its ability to recognize a single language.

Improving the quality of the audio dataset directly impacts the performance in dealing with code-switching scenarios. Our Whisper-MCE-Small model achieves a 14.28\% MER in code-switching scenarios, which is 35.13\% lower than the original model. The Whisper-MCE-Large model achieves a 12.61\% MER, which is 14.22\% lower than the original model in pure Cantonese scenarios, leading to impressive zero-shot performance.

In this Speech-to-Text application scenario, we set $\alpha=0.4$, $\beta=0.3$, and $\gamma=0.3$. Whisper-MCE-Small achieved a 90.91 FAL score in the code-switching scenario and an 87.71 FAL score in the pure Cantonese scenario, which reflects the best performance among all the models. Traditionally, larger models usually achieve higher scores, but according to FAL, Whisper-MCE-Small scored 14.1 points higher than Whisper-MCE-Large in code-switching scenarios and 8.99 points higher than Whisper-MCE-Large in pure Cantonese scenarios. The results of ablation experiments are presented in Table \ref{mix and single} and Table \ref{diff asr model}.

\subsection{Analysis of Code-switching Misrecognition}
\begin{CJK}{UTF8}{bsmi}
Table \ref{test result} presents a comparison of recognition results between the fine-tuned Whisper-MCE-small, the pre-fine-tuned Whisper-Large-V2 and Whisper-small. The content of the sentence consisted of loanwords in Cantonese and their corresponding English words. Loanwords in Cantonese are phonetically borrowed from English, such as the pronunciation of 'tips' in English being very close to the Cantonese word '貼士'. However, Whisper-small misrecognized the Cantonese word '貼士' as the English word 'Tipsy', and similar pronunciation errors occurred for other loanwords.
\end{CJK}

\section{Conclusion}
The MCE dataset addresses a notable gap in the research community and serves as a valuable resource for enhancing and developing models capable of managing code-switching scenarios, an area that has been largely neglected in previous research. Furthermore, our comprehensive evaluation metric aims to overcome the limitations of current evaluation metrics, thereby making a significant contribution to the advancement of the field.


\begin{thebibliography}{00}

\bibitem{leung2001hkcac}
Man-Tak Leung and Sam-Po Law,
\newblock ``Hkcac: the hong kong cantonese adult language corpus,''
\newblock {\em International journal of corpus linguistics}, vol. 6, no. 2, pp.
  305--325, 2001.

\bibitem{winterstein2020cantomap}
Gr{\'e}goire Winterstein, Carmen Tang, and Regine Lai,
\newblock ``Cantomap: a hong kong cantonese maptask corpus,''
\newblock in {\em Proceedings of the Twelfth Language Resources and Evaluation
  Conference}, 2020, pp. 2906--2913.

\bibitem{ardila2019common}
Rosana Ardila, Megan Branson, Kelly Davis, Michael Henretty, Michael Kohler,
  Josh Meyer, Reuben Morais, Lindsay Saunders, Francis~M Tyers, and Gregor
  Weber,
\newblock ``Common voice: A massively-multilingual speech corpus,''
\newblock {\em arXiv preprint arXiv:1912.06670}, 2019.

\bibitem{yu2022automatic}
Tiezheng Yu, Rita Frieske, Peng Xu, Samuel Cahyawijaya, Cheuk Tung~Shadow Yiu,
  Holy Lovenia, Wenliang Dai, Elham~J Barezi, Qifeng Chen, Xiaojuan Ma, et~al.,
\newblock ``Automatic speech recognition datasets in cantonese: A survey and
  new dataset,''
\newblock {\em arXiv preprint arXiv:2201.02419}, 2022.

\bibitem{gardner2009code}
Penelope Gardner-Chloros,
\newblock {\em Code-switching},
\newblock Cambridge university press, 2009.


\bibitem{kang2011mandarin}
Moonyoung Kang, Tim Ng, and Long Nguyen,
\newblock ``Mandarin word-character hybrid-input neural network language
  model.,''
\newblock in {\em INTERSPEECH}, 2011, pp. 625--628.

\bibitem{wang2003word}
Ye-Yi Wang, Alex Acero, and Ciprian Chelba,
\newblock ``Is word error rate a good indicator for spoken language
  understanding accuracy,''
\newblock in {\em 2003 IEEE workshop on automatic speech recognition and
  understanding (IEEE Cat. No. 03EX721)}. IEEE, 2003, pp. 577--582.

\bibitem{wang2022self}
Yizhong Wang, Yeganeh Kordi, Swaroop Mishra, Alisa Liu, Noah~A Smith, Daniel
  Khashabi, and Hannaneh Hajishirzi,
\newblock ``Self-instruct: Aligning language model with self generated
  instructions,''
\newblock {\em arXiv preprint arXiv:2212.10560}, 2022.

\bibitem{shen2024hugginggpt}
Yongliang Shen, Kaitao Song, Xu~Tan, Dongsheng Li, Weiming Lu, and Yueting
  Zhuang,
\newblock ``Hugginggpt: Solving ai tasks with chatgpt and its friends in
  hugging face,''
\newblock {\em Advances in Neural Information Processing Systems}, vol. 36,
  2024.

\bibitem{wu2023autogen}
Qingyun Wu, Gagan Bansal, Jieyu Zhang, Yiran Wu, Shaokun Zhang, Erkang Zhu,
  Beibin Li, Li~Jiang, Xiaoyun Zhang, and Chi Wang,
\newblock ``Autogen: Enabling next-gen llm applications via multi-agent
  conversation framework,''
\newblock {\em arXiv preprint arXiv:2308.08155}, 2023.

\bibitem{li2000cantonese}
David~CS Li,
\newblock ``Cantonese-english code-switching research in hong kong: A y2k
  review,''
\newblock {\em World Englishes}, vol. 19, no. 3, pp. 305--322, 2000.

\bibitem{li2008understanding}
David Li,
\newblock ``Understanding mixed code and classroom code-switching: Myths and
  realities.,''
\newblock {\em New Horizons in Education}, vol. 56, no. 3, pp. 75--87, 2008.

\bibitem{radford2023robust}
Alec Radford, Jong~Wook Kim, Tao Xu, Greg Brockman, Christine McLeavey, and
  Ilya Sutskever,
\newblock ``Robust speech recognition via large-scale weak supervision,''
\newblock in {\em International Conference on Machine Learning}. PMLR, 2023,
  pp. 28492--28518.

\bibitem{li2004analysis}
Yujia Li, Tan Lee, and Yao Qian,
\newblock ``Analysis and modeling of f0 contours for cantonese
  text-to-speech,''
\newblock {\em ACM Transactions on Asian Language Information Processing
  (TALIP)}, vol. 3, no. 3, pp. 169--180, 2004.

\bibitem{park2023generative}
Joon~Sung Park, Joseph O'Brien, Carrie~Jun Cai, Meredith~Ringel Morris, Percy
  Liang, and Michael~S Bernstein,
\newblock ``Generative agents: Interactive simulacra of human behavior,''
\newblock in {\em Proceedings of the 36th annual acm symposium on user
  interface software and technology}, 2023, pp. 1--22.

\bibitem{zhang2022bigssl}
Yu~Zhang, Daniel~S Park, Wei Han, James Qin, Anmol Gulati, Joel Shor, Aren
  Jansen, Yuanzhong Xu, Yanping Huang, Shibo Wang, et~al.,
\newblock ``Bigssl: Exploring the frontier of large-scale semi-supervised
  learning for automatic speech recognition,''
\newblock {\em IEEE Journal of Selected Topics in Signal Processing}, vol. 16,
  no. 6, pp. 1519--1532, 2022.

\bibitem{zheng2023judging}
Lianmin Zheng, Wei-Lin Chiang, Ying Sheng, Siyuan Zhuang, Zhanghao Wu, Yonghao
  Zhuang, Zi~Lin, Zhuohan Li, Dacheng Li, Eric Xing, et~al.,
\newblock ``Judging llm-as-a-judge with mt-bench and chatbot arena,''
\newblock {\em arXiv preprint arXiv:2306.05685}, 2023.

\bibitem{kuchaiev2019nemo}
Oleksii Kuchaiev, Jason Li, Huyen Nguyen, Oleksii Hrinchuk, Ryan Leary, Boris
  Ginsburg, Samuel Kriman, Stanislav Beliaev, Vitaly Lavrukhin, Jack Cook,
  et~al.,
\newblock ``Nemo: a toolkit for building ai applications using neural
  modules,''
\newblock {\em arXiv preprint arXiv:1909.09577}, 2019.

\bibitem{yu2020dual}
Jiahui Yu, Wei Han, Anmol Gulati, Chung-Cheng Chiu, Bo~Li, Tara~N Sainath,
  Yonghui Wu, and Ruoming Pang,
\newblock ``Dual-mode asr: Unify and improve streaming asr with full-context
  modeling,''
\newblock {\em arXiv preprint arXiv:2010.06030}, 2020.

\bibitem{chen2021developing}
Xie Chen, Yu~Wu, Zhenghao Wang, Shujie Liu, and Jinyu Li,
\newblock ``Developing real-time streaming transformer transducer for speech
  recognition on large-scale dataset,''
\newblock in {\em ICASSP 2021-2021 IEEE International Conference on Acoustics,
  Speech and Signal Processing (ICASSP)}. IEEE, 2021, pp. 5904--5908.

\end{thebibliography}
\end{document}